\documentclass[conference]{IEEEtran}
\IEEEoverridecommandlockouts
\usepackage{amsmath,amsfonts}
\usepackage{array}
\usepackage{textcomp}
\usepackage{stfloats}
\usepackage{url}
\usepackage{verbatim}
\usepackage{graphicx}
\usepackage{color}
\usepackage{svg}
\usepackage{pdfpages}

\usepackage[linesnumbered,ruled,vlined]{algorithm2e}
\RequirePackage{letltxmacro}
\LetLtxMacro{\LaTeXtextbf}{\textbf}
\LetLtxMacro{\textbf}{\LaTeXtextbf}
\usepackage{cite}
\usepackage{amsmath,amssymb,amsfonts}
\usepackage{graphicx}
\usepackage{textcomp}
\usepackage{caption}
\usepackage{subcaption}
\usepackage[bookmarks=false]{hyperref} 
\usepackage{algpseudocode}
\usepackage{array}
\usepackage{enumitem,lipsum}
\setlist[itemize,enumerate]{leftmargin=*}
\usepackage{etoolbox}

\def\BibTeX{{\rm B\kern-.05em{\sc i\kern-.025em b}\kern-.08em
    T\kern-.1667em\lower.7ex\hbox{E}\kern-.125emX}}
\usepackage{tcolorbox}
\tcbuselibrary{breakable}

\newcommand{\ie}{\textit{i.e.}}

\begin{document}

\title{RollupTheCrowd: Leveraging ZkRollups for a Scalable and Privacy-Preserving Reputation-based Crowdsourcing Platform}

\author{
\IEEEauthorblockN{Ahmed Mounsf Rafik Bendada, Mouhamed Amine Bouchiha, Mourad Rabah, Yacine Ghamri-Doudane}
\IEEEauthorblockA{ L3i - La Rochelle University, La Rochelle, France \\
\{ahmed.bendada, mouhamed.bouchiha, mourad.rabah, yacine.ghamri\}@univ-lr.fr}}

\maketitle

\begin{abstract}

Current blockchain-based reputation solutions for crowdsourcing fail to tackle the challenge of ensuring both efficiency and privacy without compromising the scalability of the blockchain. Developing an effective, transparent, and privacy-preserving reputation model necessitates on-chain implementation using smart contracts. However, managing task evaluation and reputation updates alongside crowdsourcing transactions on-chain substantially strains system scalability and performance. This paper introduces RollupTheCrowd, a novel blockchain-powered crowdsourcing framework that leverages zkRollups to enhance system scalability while protecting user privacy. Our framework includes an effective and privacy-preserving reputation model that gauges workers' trustworthiness by assessing their crowdsourcing interactions. To alleviate the load on our blockchain, we employ an off-chain storage scheme, optimizing RollupTheCrowd's performance. Utilizing smart contracts and zero-knowledge proofs, our Rollup layer achieves a significant 20x reduction in gas consumption. To prove the feasibility of the proposed framework, we developed a proof-of-concept implementation using cutting-edge tools. The experimental results presented in this paper demonstrate the effectiveness and scalability of RollupTheCrowd, validating its potential for real-world application scenarios.
\end{abstract}

\begin{IEEEkeywords}
Blockchain, Decentralized Reputation, Crowdsourcing, Privacy, zkRollups
\end{IEEEkeywords}

\begin{tcolorbox}[breakable,boxrule=1pt,colframe=black,colback=white]
\scriptsize Paper accepted at IEEE 48th Annual Computers, Software, and Applications Conference (COMPSAC) IEEE, Osaka, Japan (2024).
\end{tcolorbox}

\maketitle

\section{Introduction}\label{sec:introduction}

\IEEEPARstart{A}s a result of the growth of the Internet and the prevalence of mobile devices, crowdsourcing become increasingly popular, and the platforms facilitating this approach have experienced a significant surge in usage and recognition. The term crowdsourcing was first introduced by Jeff in 2006 \cite{b1}, It refers to a collaborative approach that delegates tasks, problems, or ideas to a broad collective. Mobile crowdsourcing, on the other hand, utilizes mobile devices like smartphones to perform the tasks. Crowdsensing, relatedly, gathers data from numerous individuals via IoT devices, such as smartphones and integrated sensors, to understand the physical environment. A prominent illustration of crowdsourcing is exemplified by Wikipedia\footnote{\href{https://en.wikipedia.org}{wikipedia.org}}

\quad Most current crowdsourcing platforms, such as Fiverr\footnote{\href{https://www.fiverr.com/}{fiverr.com}}, are centralized, which raises concerns about privacy, security, and transparency. Centralization implies a concentration of control, wherein a single authority holds sway over operations and data. This concentration raises privacy worries as user information and activities may be more susceptible to breaches or misuse. Security becomes a pressing issue due to the vulnerability of a central point of access, potentially exposing the platform to various risks and threats. Moreover, the lack of transparency in decision-making or data handling within such centralized platforms can lead to ambiguity, eroding users' trust and understanding of how their information is managed and utilized. In reputation-centric crowdsourcing systems, transparency gains even greater importance as users need to know how their reputation scores are maintained. More precisely, they seek the ability to track and verify updates to their scores at any given moment.

\quad In the last few years, numerous efforts have arisen to leverage blockchain technology in addressing these issues \cite{b7,b9}. The decentralization, transparency, and efficiency brought by blockchain are clearly what we always hoped for to build effective trustless reputation systems for crowdsourcing or any real-world application. However, transparent and effective blockchain-based reputation management requires the reputation model to be implemented on-chain often using smart contracts to enhance trust and achieve accountability. Unfortunately, in this situation, the blockchain is required to handle additional transactions such as task evaluation and overall reputation updates. This added workload significantly impacts both the scalability and performance of the system, leading to heightened gas costs, prolonged processing times, and increased time overhead. Consequently, addressing these challenges becomes imperative as they stand as substantial barriers to the practical implementation of this solution in real-world situations.

Motivated by the above challenges, our contribution presented in this paper covers the following points:
\begin{itemize}
\item A blockchain-powered fully decentralized platform to manage the entire reputation-based crowdsourcing process.
\item RollupTheCrowd leverages zkRollups (Layer-2) for empowering scalability by alleviating the burden on the mainchain (Layer-1).
\item A privacy-preserving reputation model adaptable to diverse crowdsourcing scenarios, and resilient against common reputation attacks.
\item A secure and robust crowdsourcing smart contracts automation using Decentralized Oracle Networks (DON).
\item The proposed solution is supported by a concrete proof of concept implemented using emerging technologies. RollupTheCrowd code is available on Github\footnote{\href{https://github.com/0xmoncif213/RollupTheCrowd}{https://github.com/0xmoncif213/RollupTheCrowd}}
\item Both the analytical and experimental evaluations validate the efficiency and scalability of our framework.
\end{itemize}

The remaining organization of this paper is as follows. First, the preliminaries are introduced in Section \ref{sec:preliminary}. The existing related literature is summarized in Section \ref{sec:relatedWork}.  Section \ref{sec:proposedModel} presents the overall framework proposed in this paper and describes the designed crowdsourcing scheme. Section \ref{sec:reputationFramework} details the proposed reputation model. Section \ref{sec:proofOfConcept} is devoted to the proof of concept presentation and its performance analysis. Finally, Section \ref{sec:conclusion} concludes our paper and discusses future work.

\section{Preliminary} \label{sec:preliminary}

\begin{itemize}

\item \textit{InterPlanetary File System (IPFS):} is a peer-to-peer distributed file system that seeks to connect all computing devices with the same system of files. IPFS aims to replace the traditional centralized model of the Internet with a decentralized and more resilient system. It uses a Distributed Hash Table (DHT) to address content by its hash, making it efficient, secure, and resistant to censorship. IPFS is often utilized for Decentralized Applications (DApps) and to build a more robust and accessible Internet infrastructure \cite{b13}.
\item \textit{Decentralized Oracle Networks (DONs):} are decentralized systems that facilitate the retrieval and delivery of external data to smart contracts in a decentralized and trustless manner. These oracles serve as bridges between blockchain networks and real-world data sources (APIs, external systems...). While storing all the data on-chain is non-suitable, oracles solve the problem of the inability of smart contracts to access data that are not already stored on-chain which can be a limiting factor for many application scenarios such as that of multi-party business processes \cite{b14}.
\item \textit{Rollups:} A Layer-2 (L2) scaling solution that offers a method to streamline the validation of transactions, cutting down on the resources and time needed by minimizing the data each node must process. This optimization is achieved through a secondary layer network involving actors who handle transactions off the primary chain. Subsequently, the transaction data is consolidated into batches and broadcasted onto the Layer-1 (L1) blockchain. There exist two types of Rollups, Optimistic and Zero-Knowledge (zk) Rollups. Optimistic rollups assume that transactions are valid and no computation for verification is done by default to significantly improve scalability. In zk Rollups, on the other hand, each batch contains a cryptographic proof. Calculating the proofs is complex, but checking them on the mainchain is fast \cite{b15}. 
\end{itemize}

\section{Related work} \label{sec:relatedWork}
\quad Having described the preliminaries of this work, Let us now review previous efforts that have contributed to the decentralization of crowdsourcing platforms using blockchain. 

\quad zkCrowd \cite{b7} presents a hybrid blockchain crowdsourcing platform with two ledgers. It combines Delegated Proof of Stake (DPoS) and Practical Byzantine Fault Tolerance (PBFT). These consensus protocols are chosen for their good performance, but the dual use of these protocols introduces complexity and vulnerability into the design of the hybrid blockchain without sufficient feasibility analysis. CHChain \cite{b16} framework also proposes a hybrid structure and uses a Reputation-based PBFT consensus scheme to improve system throughput. However, the feedback-based reputation model is vulnerable to bad collusion attacks, raising concerns about the security of the whole system. RC-CHAIN \cite{b17} focuses on vehicular data sharing, using a consortium blockchain. However, it introduces centralization through Roadside Units (RSUs), acting as intermediaries. In \cite{b18,b19}, supervised blockchain architectures are adopted for mobile crowdsourcing(sensing), introducing centralization concerns with a Key Distribution Center (KDC) and Task Distribution Center (TDC). 

\begin{figure}[t]
\centering
\includegraphics[width=0.8 \linewidth]{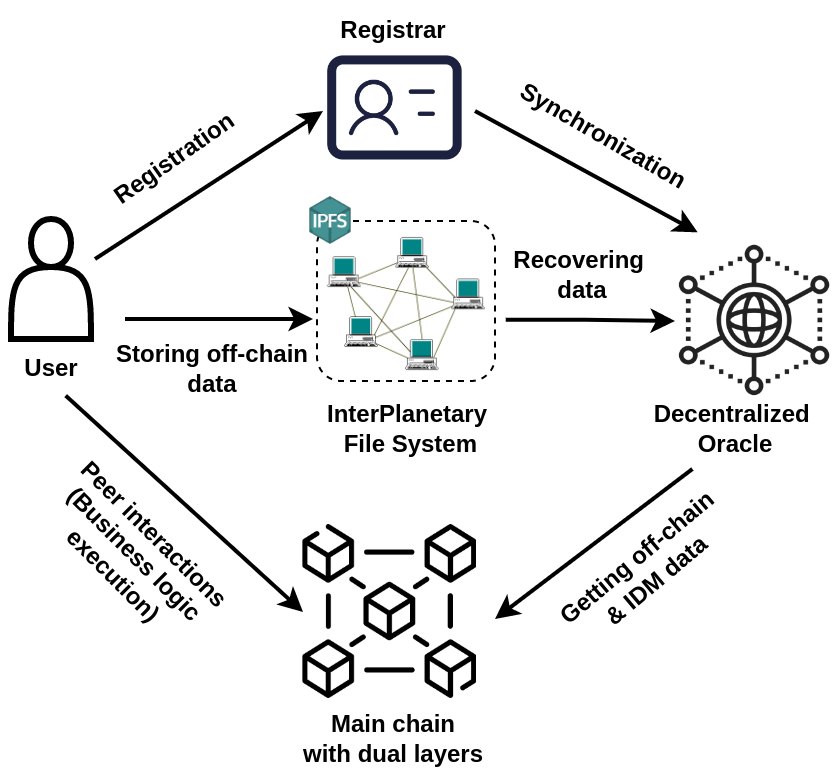}
\caption{High-Level System Architecture}
\label{fig:high-level-architecture}
\end{figure}

\quad In \cite{b21}, a decentralized reputation system for E-commerce stores content on IPFS, addressing content volume concerns without explicitly delving into other issues such as identity management and scalability. It also groups evaluations and considers transaction magnitude, interaction time, and historical reputation scores which lead to linkability and privacy exposure. RBT \cite{b22} tailors reputation assessment based on individual roles, raising re-entry attack concerns. ExCrowd \cite{b23} addresses challenges for newbies with an exploration approach through linear regression and decision tree algorithms, these algorithms can be computationally intensive and lack flexibility once are deployed. In \cite{b18}, a reputation model focuses on data reliability for the crowdsensing use case, noting potential issues with storing a high volume of sensed data. Ensuring scalability is fundamental in the design of blockchain solutions. Nevertheless, certain previous studies \cite{b9, b21, b23} tend to disregard the challenges associated with scalability. Meanwhile, alternative methods employing rapid consensus protocols to scale the system \cite{b7, b22} or resorting to centralization for enhanced performance \cite{b16, b18} have proven ineffective and, at times, insufficiently secure. L1 scaling solutions are essential but may not provide a comprehensive solution. 

\quad In summary, balancing decentralization, privacy preservation, and scalability is vital for building feasible and robust blockchain-based crowdsourcing solutions. The aforementioned studies present numerous limitations that prevent their widespread application. Therefore, to overcome these issues, this paper introduces RollupTheCrowd, a scalable, privacy-preserving, and fully decentralized reputation-based crowdsourcing framework.

\begin{figure}[t]
 \centering
\includegraphics[width=1 \linewidth]{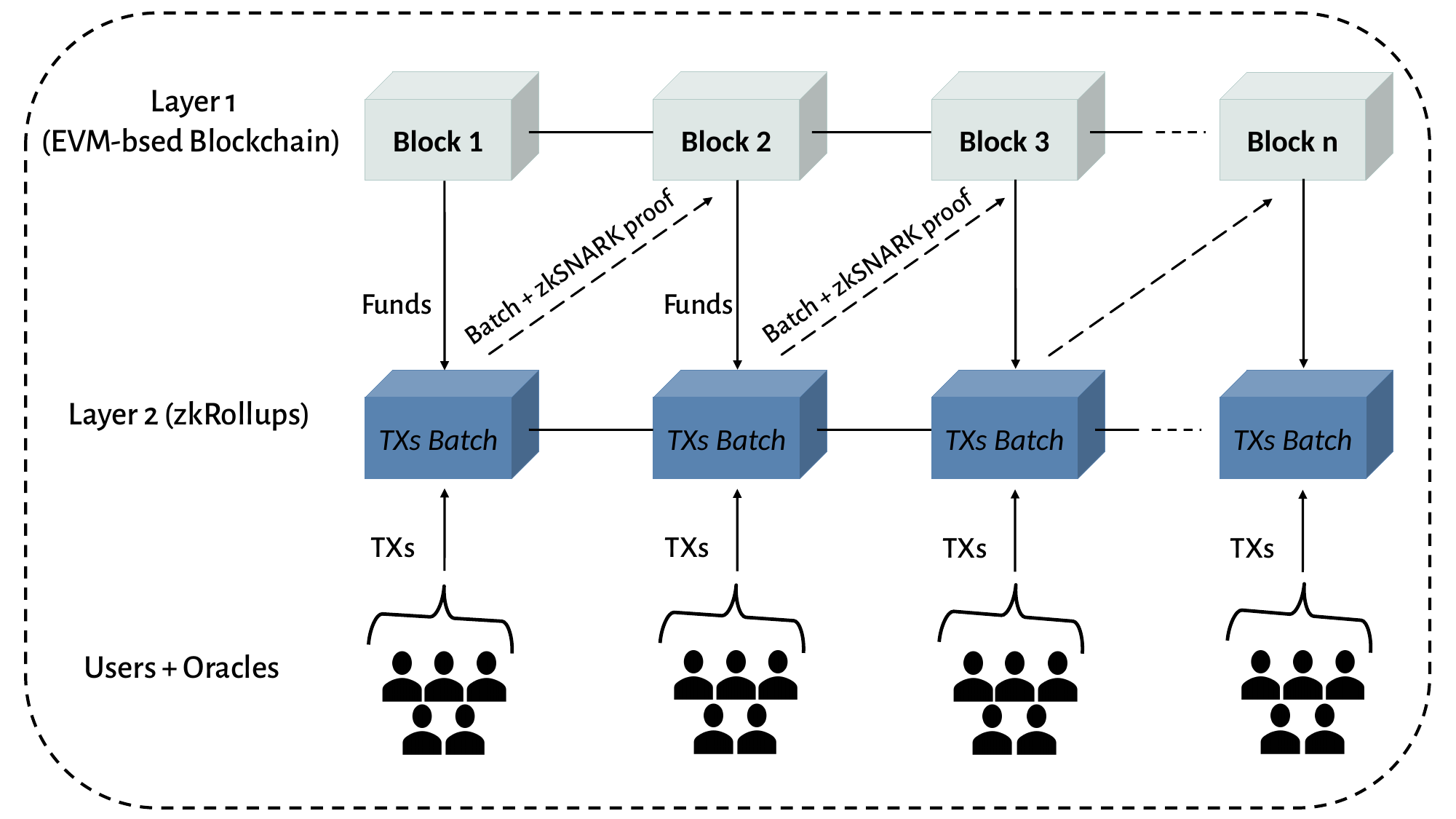}
\caption{Dual Layers Workflow in RollupTheCrowd}
\label{fig:dual-layer}
\end{figure}

\section{RollupTheCrowd Framework} \label{sec:proposedModel} 

\quad After exploring related studies and the challenges identified in current solutions, in this section, we will present our framework. We begin by describing the complete architecture of the system and then detailing all the components of the proposed solution.   

\subsection{System Architecture}

\quad Figure \ref{fig:high-level-architecture} shows the proposed architecture for RollupTheCrowd. It has four components.

\subsubsection{Main Ledger with Dual Layers for Enhanced Scalability and Cost Efficiency}
The main ledger is the central element of our crowdsourcing platform, featuring a dual-layer structure as shown in Figure \ref{fig:dual-layer}.  The first layer operates as a traditional blockchain network, employing a Proof of Authority (PoA) consensus, ensuring the security and scalability of the mainchain (L1). The second layer employs zero-knowledge (zk) Rollups solution to enhance scalability. Instead of processing each transaction on the main chain, a batch of transactions is processed and validated off-chain (on L2), by the aggregator. It then publishes the new state root, compressed transaction data, and proof of validity on the main chain. This proof of validity ensures the computation made to execute the transactions was correct. zkRollups inherent security from L1 and upholds privacy by design, which makes them the perfect solution for our underlying crowdsourcing system. This design not only guarantees transparency for all participants but also upholds the decentralized nature of the system, reducing congestion and lowering fees. At the same time, the security guarantees of the main blockchain are preserved through cryptographic proofs.

\subsubsection{A Decentralized Registrar for a Secure Identity Management} Complementing the prowess of the main ledger, we introduce a Registrar Ledger, responsible for identity management in our decentralized ecosystem. This ledger serves as an identity management entity such as CanDID \cite{b10}, It oﬀers a secure environment where users can assert their identities, without sending any information but only proofs, The role of the registrar in our protocol can be assumed by the CanDID committee, a decentralized set of nodes, which performs deduplication (identity uniqueness) in a privacy-preserving way.

\subsubsection{Dual blockchain ledgers with IPFS Integration} The main ledgers in our design are coupled with an InterPlanetary File System (IPFS), providing an efficient solution to the storage challenges inherent in traditional blockchain-based crowdsourcing systems. By oﬄoading substantial data oﬀ-chain to IPFS, the transaction times and costs within RollupTheCrowd are optimized.   
\begin{figure}[t]
\centering
\includegraphics[width=1 \linewidth]{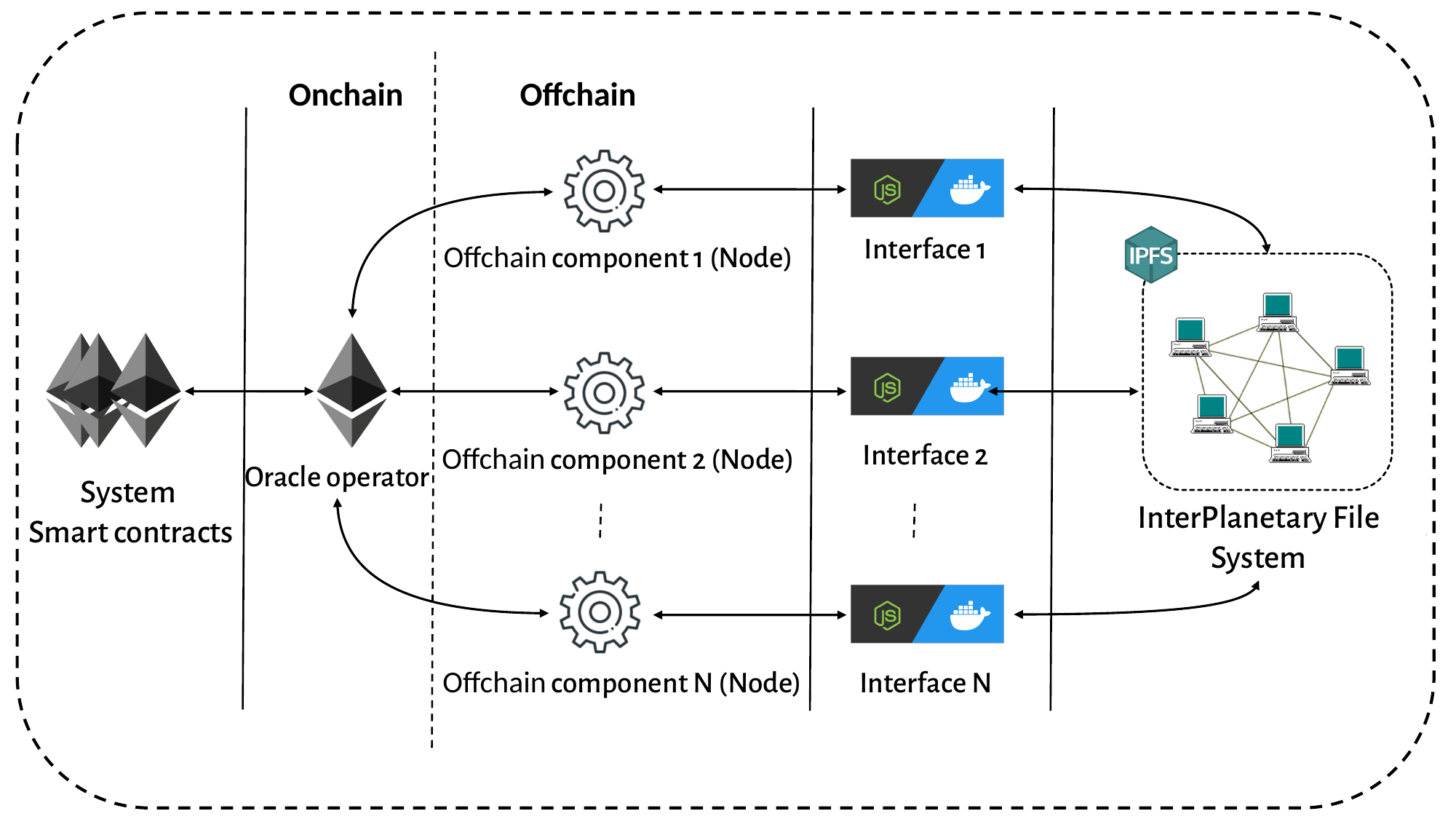}
\caption{Decentralized Oracles in RollupTheCrowd}
\label{oraclefig}
\end{figure}
\subsubsection{Interoperability and Synchronization}
An essential facet of our system architecture is the seamless interaction between these two ledgers. Smart contracts deployed on the main ledger can be triggered by authorized users to execute operations, This can be achieved using the inter-chain decentralized oracles, ensuring that the data shared between the ledgers is accurate, tamper-proof, and auditable. The inclusion of IPFS and the oﬀ-chain storage of business logic data necessitates a reliable mechanism for data synchronization and retrieval. This is where decentralized oracles excel, ensuring that data from IPFS can be efficiently utilized on the main blockchain without compromising security or decentralization. Figure \ref{oraclefig} explains how these decentralized oracles serve as a bridge between on-chain and off-chain sides \cite{b31}.

\subsection{RollupTheCrowd Modules}

\quad Within RollupTheCrowd, blockchain nodes at L1 can independently transmit, verify, and store data within the network. They are also responsible for validating transactions/blocks and reaching a consensus on the state of the blockchain. Below, we list the functional modules we developed and integrated into each blockchain node.

\begin{enumerate}
    \item \textbf{Oracle Operator Module:} An Oracle operator module is a smart contract within a blockchain ecosystem that acts as an intermediary or bridge between the blockchain and external data sources. Its primary purpose is to fetch, verify, and provide oﬀ-chain data to on-chain smart contracts, enabling them to interact with real-world information.

\item \textbf{Access Management Module:} is a smart contract that manages access
control and permissions within our Decentralized application (DApp). It is a common pattern used to control who can perform certain actions or access speciﬁc functionalities within the application. The primary objective of this is to ensure that only authorized users or addresses are allowed to execute speciﬁc operations or access sensitive data.

\item \textbf{Business logic Module:} is the smart contract that facilitates the management
of crowdsourcing operations within our system. It enables users to create, submit, and complete tasks in a decentralized manner. It implements the complete business logic behind the crowdsourcing scenario (details will follow).

\item \textbf{Reputation Module:} is the component responsible for managing reputation scores in the system. It implements our proposed privacy-preserving reputation model (details in Sec. \ref{sec:reputationFramework}). 

\end{enumerate}

\quad It's important to point out that with the integration of this reputation module into the framework, we have the option of employing reputation-centric consensus \cite{guru,b24}. This alternative offers better scalability and fairness than PoA and PoS respectively. We will further explore this aspect in an extended version of this work.

\subsection{Smart Contracts Design} \label{subsec:smartcontract}

\quad The entire business logic of RollupTheCrowd is implemented using smart contracts. 
We employ three primary entities in our crowdsourcing process: Requesters, Workers, and Evaluators. Smart contracts (SCs) in RollupTheCrowd, are designed to provide transparency and accountability among these entities by managing both reputation and crowdsourcing tasks on-chain. In the following, we detail the core functions executed through smart contracts in our system.

\begin{figure}[t]
\centering
\includegraphics[width=0.8 \linewidth]{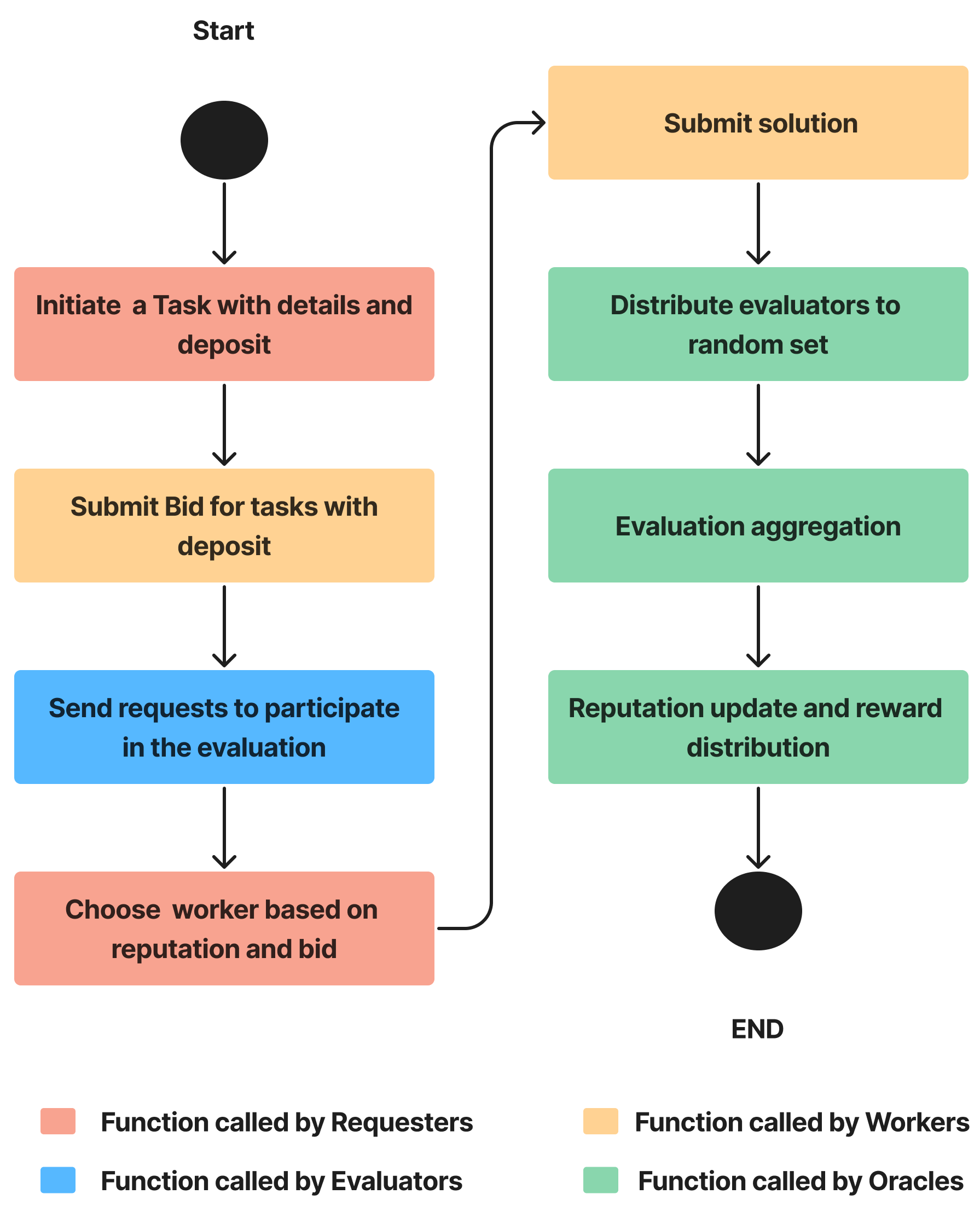}

\caption{The crowdsourcing workflow in RollupTheCrowd} 
\label{fig:workflow}
\end{figure}

\quad Figure \ref{fig:workflow} depicts the crowdsourcing workflow within our framework, illustrating SCs functions called by each entity. Deposits are locked into a SC as collateral for transactions, and any misconduct will result in penalties. Evaluators are selected before Workers, with Workers unaware of the Evaluators' identities. To ensure fair evaluations, Evaluators are randomly assigned to evaluate tasks. This setup encourages timely and high-quality contributions from Workers while maintaining the integrity of evaluations by Evaluators.

\subsubsection{Create Task}
The algorithm \ref{algo:createTask} highlights the steps of the $createTask$ function. The function initially verifies whether the caller is a registered user; if affirmative, it proceeds to store only the necessary data on-chain, as all other details have already been submitted off-chain to IPFS via the front end. Additionally, the function updates the amount to be used later in the reputation model.

\begin{algorithm}[t]
\DontPrintSemicolon
\SetAlgoNlRelativeSize{-1}

\SetKwFunction{Fassert}{Assert}
\SetKwFunction{FupdateAmount}{updateMinMaxAmounts}
\SetKwProg{Fn}{Function}{:}{}
\SetKwInOut{Input}{input}
\SetKwInOut{Output}{output}

$T$: New Task (each task $T$ has: CID, amount)

  \BlankLine

\Begin{
    \Fassert{$isUser(msg.sender) = true$} \\
    $tasks[T.CID].requester \leftarrow msg.sender$ \\
    
     $tasks[T.CID].CID \leftarrow T.CID$ \\

      $tasks[T.CID].amount \leftarrow T.amount$ \\
       \Comment{update $A_{min}$ and $A_{max}$ of our reputation model}  \\
    \FupdateAmount{$T.amount$}\\
}
\caption{Create Task}
\label{algo:createTask}
\end{algorithm}

\begin{algorithm}[t]
\DontPrintSemicolon
\SetAlgoNlRelativeSize{-1}

\SetKwFunction{Fassert}{Assert}
\SetKwFunction{FupdateAmount}{updateAmount}
\SetKwProg{Fn}{Function}{:}{}
\SetKwInOut{Input}{input}
\SetKwInOut{Output}{output}

$T$: New Task (each task $T$ has: CID, amount, bid Progress, and bid submitter) \\
$B$: Task bid (each bid $B$ has: CID) \\
$S$: Task Submission (each submission $S$ has: CID)

\Begin{
    \Fassert{$tasks[T.CID].bidSubmitter[B.CID] = msg.sender$} \\
    \Fassert{$tasks[T.CID].bidProgress[B.CID].$\\  $answer = true$} \\
    $tasks[T.CID].bidProgress[B.CID].submission \leftarrow S.CID$
    
}
\caption{Submit solution}
\label{algo:submit}
\end{algorithm}
\subsubsection{Submit Solution} The Algorithm \ref{algo:submit} implements the $submitSolution$ function, triggered by the worker upon task completion. This function first verifies if the bidder submitting the solution is indeed the assigned worker, ensuring that workers can only submit solutions within their bids. Subsequently, it checks whether the bid has been accepted by the requester, allowing only accepted workers to submit their solutions. Finally, it stores the Content Identifier (CID) of the submitted solution to IPFS.

\subsubsection{Distribute Evaluators to Random Sets}
The Algorithm \ref{algo:ditribute} implements how we distribute evaluators into random sets to achieve randomness in the evaluation process, each set is responsible for one submission. The function is called only by the oracle when there are enough evaluators for the task.
\begin{algorithm}[t]
\DontPrintSemicolon
\SetAlgoNlRelativeSize{-1}

\SetKwFunction{Fassert}{Assert}
\SetKwFunction{FupdateAmount}{updateAmount}
\SetKwFunction{Fpermutation}{permutation}

\SetKwProg{Fn}{Function}{:}{}
\SetKwInOut{Input}{input}
\SetKwInOut{Output}{output}

$T$: New Task (each task $T$ has: CID, bids , and evaluators) \\
$B$: Task bid (each bid $B$ has: CID) \\
$S$: Task Submission (each submission $S$ has: CID)

\Begin{
    $evaluators \leftarrow tasks[T.CID].evaluators $ \\
     $numSets \leftarrow tasks[T.CID].bids.length $  \\
    \For{$i \in [0,1,...,evaluators.length]$}{
        $n \leftarrow i+ (keccak256(block.timestamp)$ \\$mod (evaluator.length - i))$ \\
        \Fpermutation{$evaluators[n],evaluators[i]$} \\
        
   }
    \For{$i \in [0,1,...,evaluators.length]$}{
        $tasks[T.CID].evaluatorSets[i$ $mod$ $ numSets$] $\leftarrow evaluators[i]$ \\
        
   }
}

\caption{Distribute Evaluators To Random Sets}
\label{algo:ditribute}
\end{algorithm}

\subsubsection{Calculate New Reputation}
Upon completion of the evaluation of a specific task using corresponding measures detailed in Sec. \ref{sec:reputationFramework}, Evaluators transmit their local ratings to the Oracle. The Oracle network checks the validity of these ratings and then calculates the average scores and submits the result on-chain through the invocation of the $calculateNewRep$ function. This function computes the new reputation using the proposed reputation model, considering the task type, and then initiates the $updateReputation$ function.

\section{Reputation Modelling} \label{sec:reputationFramework}

\quad After describing the architecture of RollupTheCrowd and its main components, we will now delve into the mathematical details of the proposed reputation model.

We propose a reputation model that can be adapted to various crowdsourcing situations such as situations that involve solving complex problems, collecting data, conducting research, or harnessing collective intelligence. Those diverse scenarios can be categorized from our perspective into two principal categories: problem-solving and knowledge acquisition. In problem-solving situations, crowdsourcing initiatives focus on human-level intelligence or expertise to perform. Participants bring reasoning, problem-solving, decision-making, and learning, among other cognitive abilities that are characteristic of human intelligence. Examples of such scenarios include crowdsourcing platforms dedicated to innovation, where individuals contribute creative solutions for product development or process improvement. On the other hand, knowledge acquisition situations in crowdsourcing aim to gather a broad range of data from a diverse group of individuals or machines. This may involve the crowdsensing use case. 

\subsection{Task Evaluation}
\quad Recognizing the two types of crowdsourcing situations allows us to design targeted evaluation strategies and approaches that meet the specific needs and goals of each scenario. We define a common metric for all crowdsourcing scenarios, namely value rating. We first present this common metric and then the metrics specific to each situation.

\subsubsection{Common Metric - Value Rating} 
It is not expensive for a malicious requester to submit multiple low-cost tasks (i.e., micro-tasks) addressed to a particular worker to improperly boost their reputation. Therefore, to mitigate this type of coordinated attack and tackle the problem of unfair ratings, the rating of a task should be related to its amount $A_t$. The value rating $V_R \in [0,1]$ is computed using the following formula:
\begin{equation}
V_R = f(A_t)= \dfrac{A_t-A_{min}}{A_{max}-A_{min}} \label{valueRating}
\end{equation}

\subsubsection{Problem-Solving Tasks Metrics}
 Problem-solving tasks refer to cognitive tasks that require human-level intelligence or expertise to perform. These tasks typically involve reasoning, problem-solving, decision-making, and learning, among other cognitive abilities that are characteristic of human intelligence. Examples of human intelligence tasks include natural language understanding, logical reasoning, creativity, and social intelligence. Evaluating these tasks involves subjective judgments, which can vary from one evaluator to another. For instance, tasks that require creativity may elicit multiple valid solutions, leading to diverse ideas and approaches among individuals. To minimize conﬂicts in evaluation, we introduce objectivity and establish clear criteria during the task posting phase. By providing explicit guidelines and speciﬁcations upfront, we strive to facilitate a more structured and consistent evaluation process. This helps to ensure that evaluators have a standardized framework to assess tasks and reduce discrepancies. Within each Problem-solving task, the assessment of user submission is influenced by various factors. We deﬁne two factors to assess the Worker's submission: Effort Rating and Contextual Rating.
\begin{itemize}
   
\item \textit{Effort Rating ($E_R$):} to bring more objectivity to feedback submission, we gauge the
    user eﬀort on the task by considering the two following parameters:
    \begin{enumerate}
        \item \textit{Task completeness:} $C_t \in [0,1]$ designates the degree of completion or realization of a task or project. It is a measure of progress toward the task goal and can be computed using a deﬁned checklist by the requester. 
        \item \textit{Task Quality:} $Q_t \in [0,1]$ refers to the level of expertise or efficiency in performing a specific task. It can be calculated using a set of rubric rules defined by the requester. Rubric rules are criteria or guidelines used to evaluate the quality of an assignment. For example, for a logo design task, quality evaluation using rubric rules may include creativity and originality, relevance to brand identity, technical execution, and aesthetic appeal. Each of these metrics can be rated on a scale of one to ten. 
    \end{enumerate}
 We give requesters the freedom to determine the weighting of completeness and quality, enabling them to specify their preferences in advance. Therefore, the effort rating is computed as follows: 
 \begin{equation}  \label{effortRating}
E_R = f(C_t,Q_t)= \alpha C_t + \beta Q_t \: ; \: \alpha + \beta = 1
\end{equation}

\item \textit{Contextual Rating ($C_R$):} Worker submissions can be evaluated taking into account additional validity aspects, which may differ depending on the use case. For example, in programming contexts, considerations may relate to the success of test cases or the cleanliness of the code, enabling a more in-depth and personalized assessment to measure the quality and effectiveness of the submission.
\end{itemize}

\quad The overall rating of the problem-solving task $T_R = f(V_R,E_R,C_R)$, which is a linear combination between the three metrics. The weighting of each metric is determined by either the group or the platform operator.

\subsubsection{Knowledge Acquisition Task Metrics} 
 In this type of task, the focus is on collecting data. The gathered knowledge can be evaluated by its reliability and can be provided by IoT devices (temperature, pressure, etc.) or humans (Location, pictures, surveys). There are many existing methods for evaluating data reliability. Inspired by the method proposed in \cite{b18}, we develop a new evaluation method that enables accurate estimation of knowledge acquisition based on the reliability of acquired data.
\begin{itemize}
\item \textit{Data Distortion Rating ($D_R$):}
Data distortion represents the diﬀerence between observation and truth. We use the deviation of sensed data $V_i$ from $V_a$ to denote the degree of data distortion. $V_a$ is the ﬁnal aggregation result held by the decentralized oracle, which is considered truth data. We calculate the squared diﬀerence between $V_i$ and $V_a$, then the result is normalized as the deviation of $V_i$ from $V_a$.
\begin{equation}
     d_i= (\dfrac{V_i-V_a}{b_U-b_L})^2
    \label{dataDistortion}
\end{equation}
$b_L$ and $b_U$ represent the lower and upper bounds of the sensed data range, respectively. They are used to normalize the deviation $d_i$ (\ie, $d_i \in [0, 1]$). We calculate the data distortion metric as follows:
\begin{equation}
    D_R = {1-d_i}
\end{equation}

\begin{figure}[t]     
\centering
\includegraphics[width=1 \linewidth]{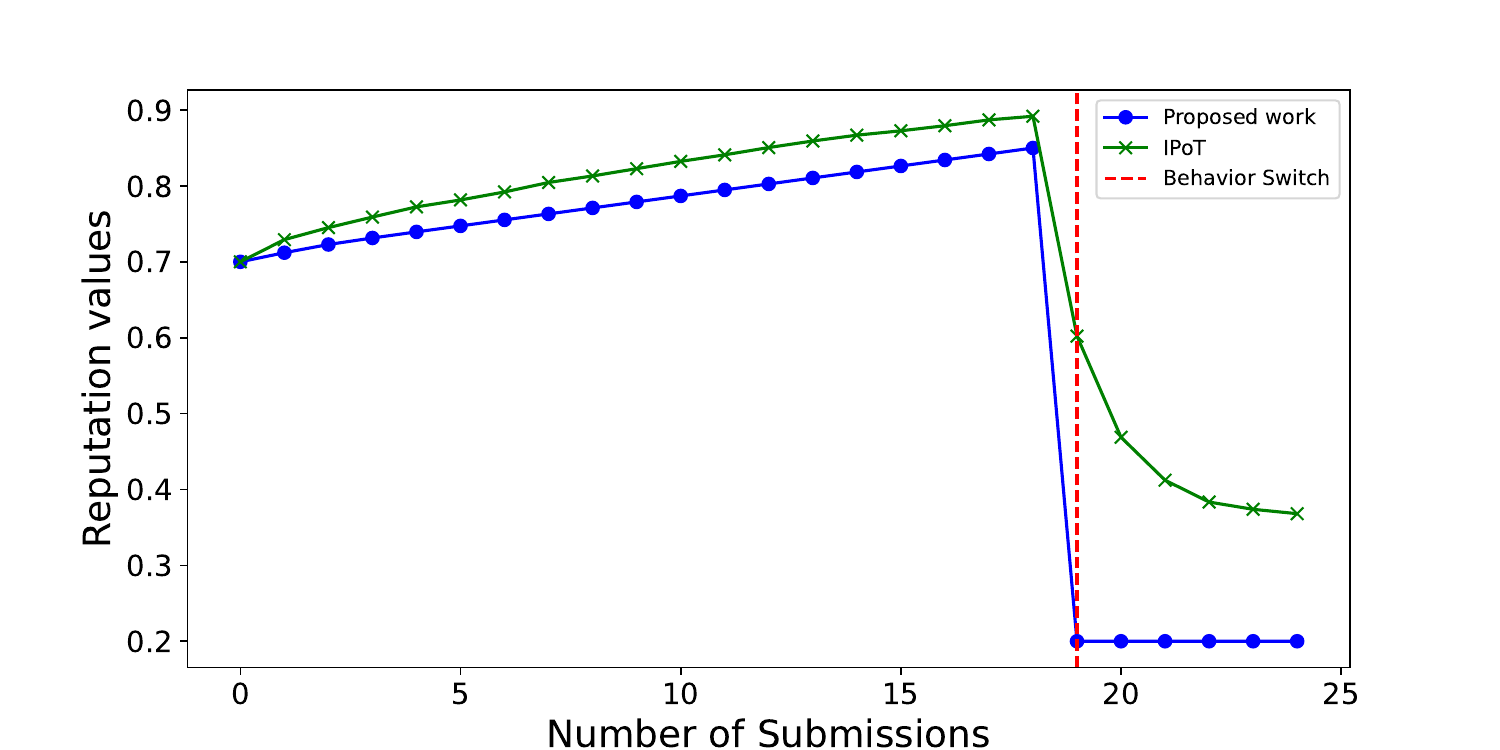}
\caption{Proposed model eﬀectiveness compared to \cite{b24}}
\label{comparison}
\end{figure}

\item \textit{Contextual Rating ($C_R$):} In addition to the data distortion rating, other contextual factors could be taken into account toward evaluating the reliability of data. For instance, the speciﬁc sensing task is strict in location and time, which means that the sensing data from the expected location might be more reliable than that from a remote location. 
\end{itemize}
\quad  Similar to the problem-solving task the overall score is $T_R = f(V_R,D_R,C_R)$ the linear combination of three ratings $V_R$, $D_R$, and $C_R$.

\subsection{Reputation Update}
\quad In our reputation calculation process, we use the task evaluation method outlined in the previous section, along with past behavior. While developing this model, we carefully considered privacy concerns. Therefore, the only data the model requires is the users' current reputation scores. In RollupTheCrowd, each new user is assigned an initial reputation score $R_{init}$. This value can be the average of the reputation scores of all existing users or another value fixed by the system operators. 

\quad Reputation is the perception that users have about an individual in the system and past behavior is a key factor in the calculation of reputation. For that, we introduced the current reputation value to calculate the new one. The reputation update process diverges based on whether the behavior exhibited is deemed good or bad. We define good work when the task value exceeds a certain threshold $T_{min}\geq R_{init}$, which we consider the critical line of trust. Otherwise, the work is categorized as bad. In the update for good behavior, we accord higher significance to the old reputation score, resulting in a reputation growth that aligns with the expected positive behavior. Contrastingly, when bad behavior occurs, we shift the focus towards the current task evaluation, imposing a stricter punishment as a consequence \cite{guru}. The general update formula is as follows:

\begin{itemize}
    \item For good behavior ($T_R \geq T_{min}$),
\end{itemize}
\begin{equation}
         R_{new}= \omega R_{old} + (1- \omega) T_R 
    \label{UpdateRep+}
\end{equation}

\begin{itemize}
    \item For bad behavior ($T_R < T_{min}$),
\end{itemize}
\begin{equation}
         R_{new}= (1- \omega) R_{old} + \omega T_R
    \label{UpdateRep-}
\end{equation}
where, $T$ is the task value, $R_{old}$ refers to the old Reputation, and $R_{new}$  is new Reputation. $\omega$ refers to the weighting function, $f_w(S) = tanh(S)$ where $S$ is the number of submissions done by the worker which leads the model to become progressively stricter as the worker engages in more tasks.

\quad The ability to respond quickly to unexpected actions is an essential feature of an eﬀective reputation model. To prove that RollupTheCrowd possesses this characteristic, we compared our model with the model employed in IPoT \cite{b24}, which uses a similar approach based on the evaluation of crowdsourcing interactions. Figure \ref{comparison} shows the variation in updates in response to positive behavior, following the user's interactions up to interaction 25, when his behavior becomes negative. As soon as a negative action is taken, the reputation score declines rapidly in both systems. However, the score decreases faster in our model. This difference reflects our system's increased ability to react to inappropriate behavior.

\section{Security Analysis}

\quad In this section, we explore the potential security vulnerabilities within RollupTheCrowd and illustrate its resilience against these threats and attacks.
\begin{itemize}
     \item \textbf{Sybil attack:} This attack involves creating multiple identities to take advantage of the reputation system. \\
    $\rightarrow$ Our system restricts the creation of multiple accounts, permitting only authentic users to join. Organizational (or consortium) admin accounts oversee system access through on-chain management using role-based access control. This protocol is established via smart contracts on-chain, granting exclusive authorization to admin accounts for user addition or removal from the system.  
    \item \textbf{Whitewashing attack: } This occurs when a dishonest worker attempts to reset their negative reputation by re-entering the system with a new identity to obtain an initial reputation score. \\
    $\rightarrow$  Users are registered through the registrar entity (a consortium of organizations). Rejoining the platform is only possible if the registrar blockchain committee grants permission. 
    \item \textbf{Collusion attacks:}
    This form of attack involves collusion between a group of workers and requesters to either lower a target worker's reputation or inflate their own. \\
    $\rightarrow$ In our system, a worker's reputation calculation is not based on the requester's feedback. It relies on evaluations conducted by a randomly selected set of evaluators. RollupTheCrowd's design prioritizes automatic evaluation metrics, making it resilient against this type of attack.
    \item \textbf{Free-riding, False-reporting:} In Free-riding workers receive rewards without making real efforts, while in False-reporting requesters try to repudiate the payment. \\
    $\rightarrow$ In our system, both of these attacks are prevented. Workers cannot receive payment without undergoing evaluation, and the reward distribution is contingent upon the outcomes of the evaluation process. Additionally, requesters are unable to repudiate payments since they are already locked within the deposit contract. Workers and requesters are obligated to make a deposit, which guarantees their commitment to the system.
    \item \textbf{Bad mouthing:} This happens when the evaluator submits an incorrect evaluation/rating. \\
    $\rightarrow$ Within our system, we address this form of attack by randomly choosing independent evaluators, as there is no advantage for evaluators in providing a false rating. Furthermore, we alleviate rating errors by calculating an average weighted score based on assessments from multiple raters. 
    
\end{itemize}

\section{Evaluation and Results} \label{sec:proofOfConcept}
\quad The main objective of RollupTheCrowd's development is to build a scalable and decentralized system capable of efficiently managing reputation and crowdsourcing tasks simultaneously. 
To validate the feasibility, scalability, and effectiveness of our proposed Framework, we developed a proof of concept for our framework. For a thorough explanation of the technical intricacies, we invite readers to explore our public GitHub repository.\footnote{\href{https://github.com/0xmoncif213/RollupTheCrowd}{https://github.com/0xmoncif213/RollupTheCrowd}}, which provides a detailed description of our implementation. This repository serves as a valuable resource for those interested in replicating and further scrutinizing our work. 

\quad This section details the experimental setup, followed by a brief overview of the technologies utilized during development and experimentation. Next, the key metric used for evaluation and the results demonstrating our system's performance for relevant benchmarks are described.

\subsection{Experimental Setup}
\quad The deployment of the proposed crowdsourcing platform and performance tests are carried out on a cluster of two servers 'HPE ProLiant XL225n Gen10 Plus' dedicated to the experimentation and evaluation of blockchain solutions. Each server is equipped with two AMD EPYC 7713 64-Core 2GHz processors and 2x256 GB RAM.

\subsection{Tools and Libraries}

\quad We implemented our smart contracts using Solidity and leveraged an Ethereum blockchain to run our network, employing Geth\footnote{\href{https://geth.ethereum.org/}{https://geth.ethereum.org/}} client. For the Oracle integration, we used Chainlink\footnote{\href{https://chain.link/} {https://chain.link/}} nodes with customized external adapters within Docker-containerized Node.js servers. \cite{b31}. For the L2 scaling solution, we incorporated zkSync\footnote{\href{https://zksync.io/}{zksync.io}} Rollups. As a comprehensive measure to validate the efficiency and scalability of the proposed solutions, we conducted thorough testing and benchmarking using the Hyperledger Caliper framework\footnote{\href{https://github.com/hyperledger/caliper-benchmarks}{https://github.com/hyperledger/caliper-benchmarks}}.

\subsection{Performance Evaluation}

\quad In our evaluation approach, we recognize the need for a fine-grained assessment of our platform’s performance. We conduct benchmarks on individual modules, including access management, business logic, pre-evaluation, and evaluation modules to achieve this. Each module has different characteristics and provides different functionality, so it is essential to tailor our benchmarking configurations (workloads) to capture the system performance associated with each module. These configurations include generating and sending transactions containing the parameters needed by each function, such as hashes, addresses, and reputation scores. For L1, the Caliper framework orchestrates workload generation and transmission, enabling preconfigured settings. Meanwhile, for L2, our approach involves designing workloads adhering to interaction standards specific to zkSync. To qualitatively assess the business logic and reputation functions, our experimental process emphasizes three essential metrics:
\begin{itemize}
    \item \textbf{Throughput:} the number of successful transactions per second (TPS).
    \item \textbf{Latency:} refers to the time diﬀerence in seconds between the submission and completion of a transaction.
    \item \textbf{Gas:} is a unit that measures the computational work required to perform operations and is influenced by the complexity of the operation, the computational steps involved, and the amount of data processed.
\end{itemize}

\begin{figure}[t]     
\centering
\includegraphics[width=1 \linewidth]{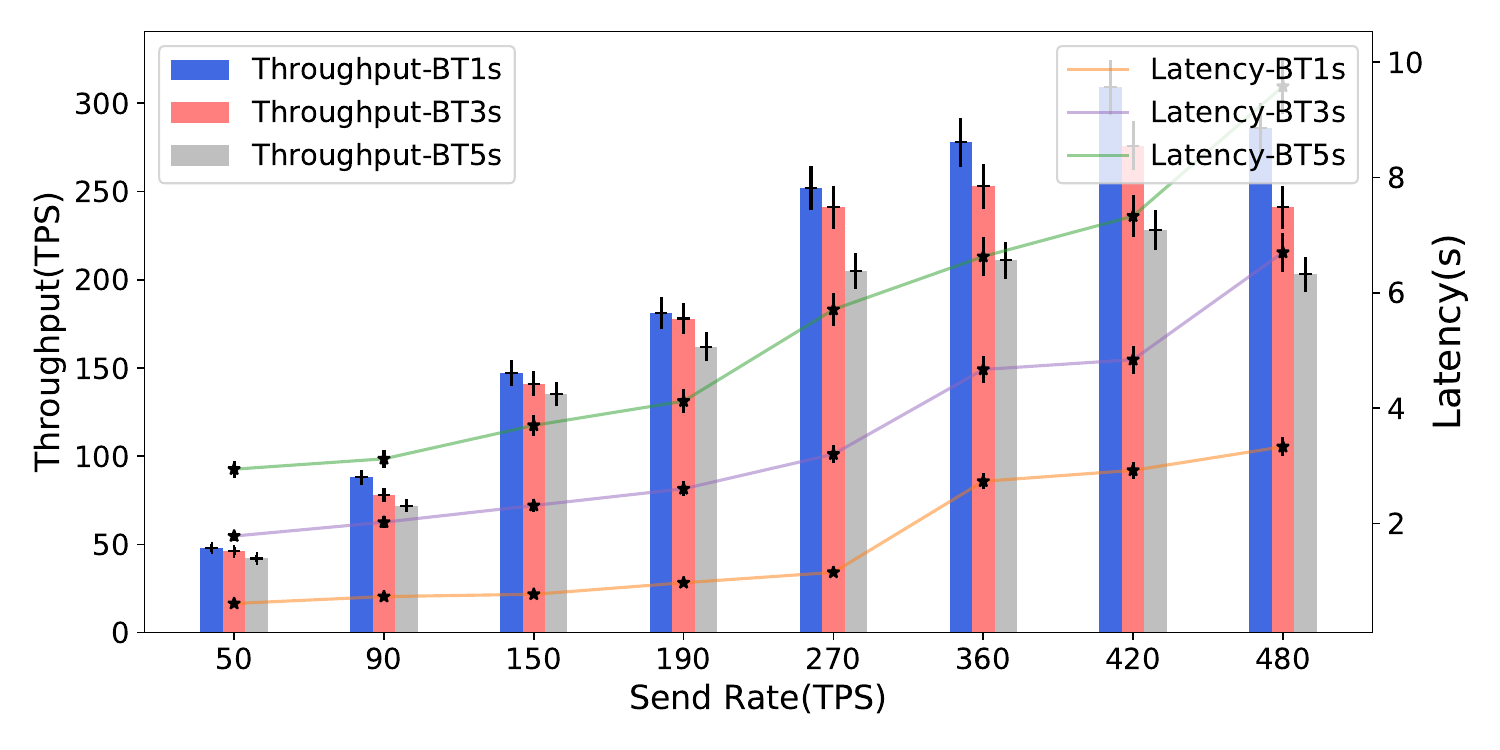}
\caption{\centering L1 Throughput and latency of CreateTask function under different Block Times = [1s,3s,5s]}
\label{createTask}
\end{figure}
\begin{figure}[t]     
\centering
\includegraphics[width=1 \linewidth]{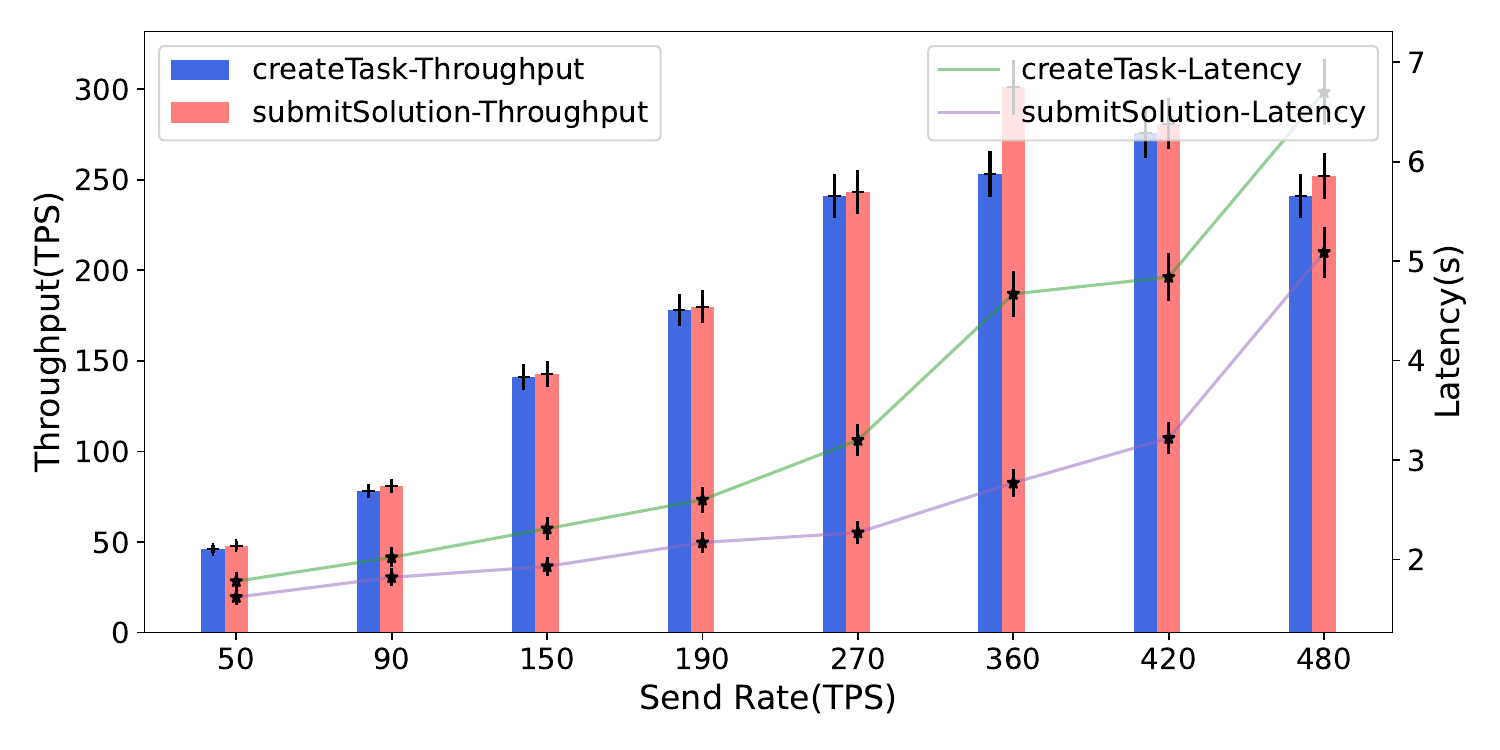}
\caption{\centering L1 Throughput and latency comparison under different  workload types for a Block Time = 3s}
\label{fig:workload-latency}
\end{figure}
\quad The results below concern the evaluation of a complete crowdsourcing problem-solving scenario, from task creation to reputation updating.

\begin{enumerate}[leftmargin=0.2cm,align=left]
\item \textit{L1 Throughput and Latency:} We begin our analysis with the performance of the mainchain. Figure \ref{createTask} illustrates the throughput and latency of the heaviest function $createTask$ in our design for different block periods (1s, 3s, 5s). Latency increases as the block period increases, which is obvious. However, even with a block duration of 5s, our approach of submitting data off-chain and storing only essential data on-chain proved to be very eﬃcient. The system achieves its best throughput of 310 (TPS) with a send rate of 420 (TPS). By changing the workload type, Figure \ref{fig:workload-latency} compares $submitSolution$ to $createTask$, the former has better throughput and latency as it requires relatively less computation on-chain.

\begin{figure*}[t]     
    \centering
    \subfloat[Create Task TX]{
        
        \includegraphics[width=0.5 \linewidth]{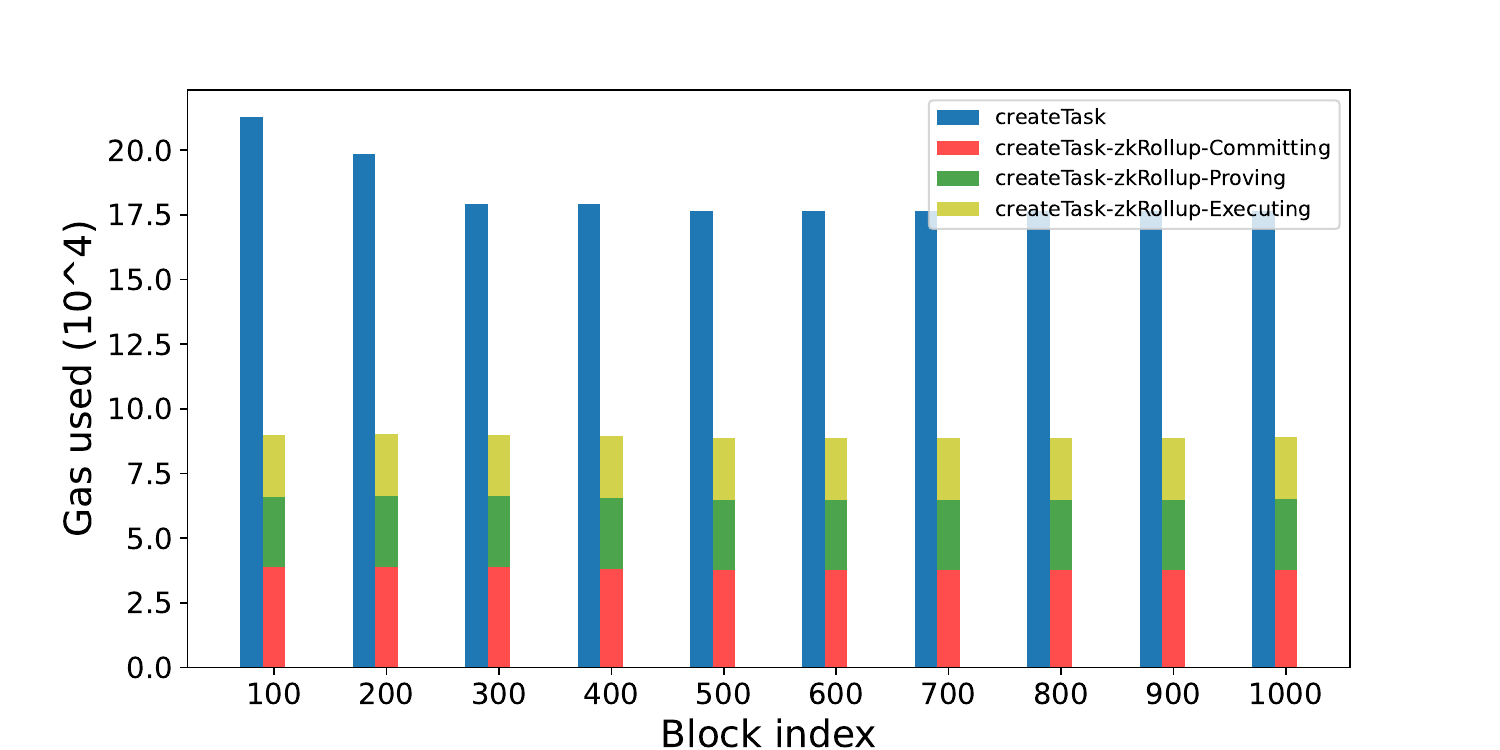}
        \label{fig8a}
   }
    \subfloat[Calculate New Reputation TX]{
        \includegraphics[width=0.5 \linewidth]{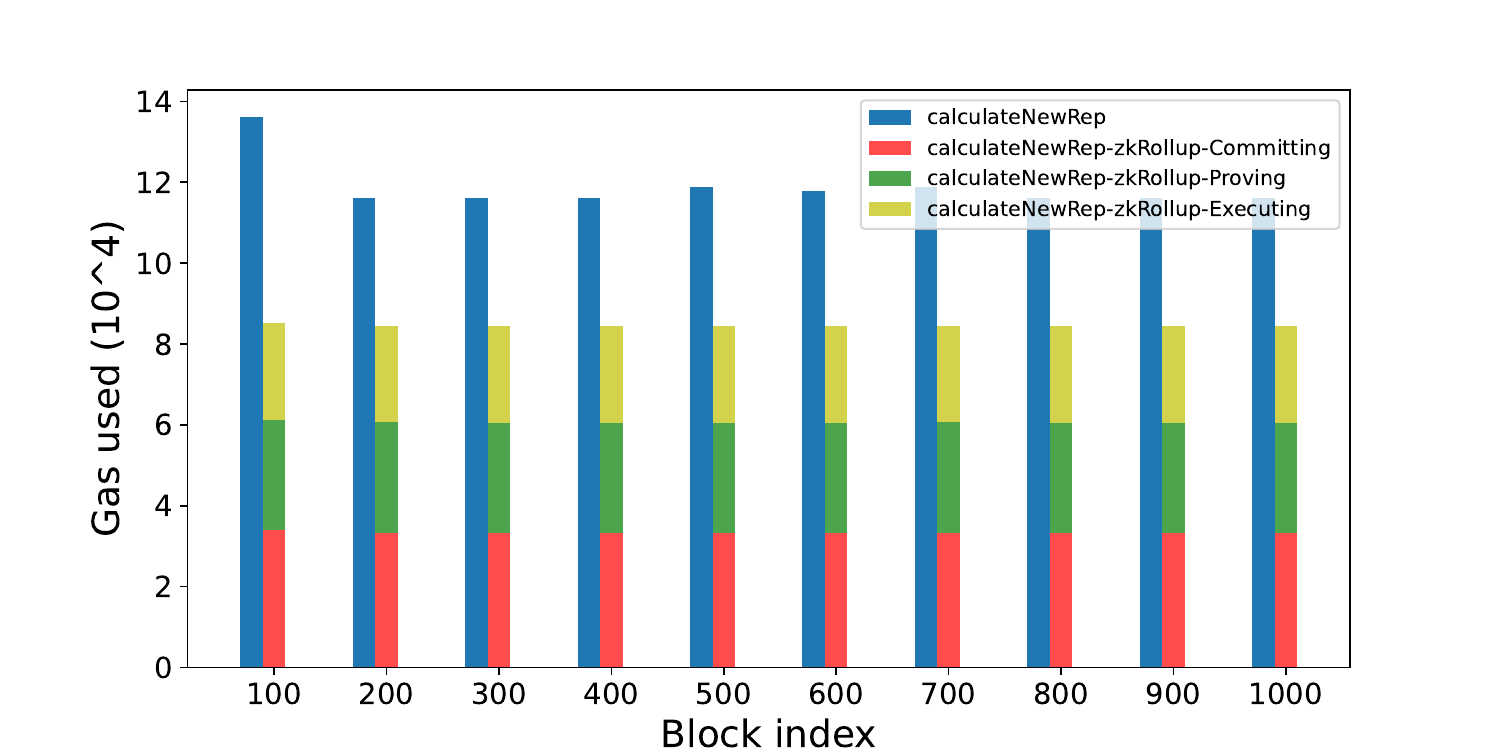}
        \label{fig8b}
   }
  
    \caption{ \centering Gas Consumption: A Comparison Between L2 and Non-L2 Implementations}
    \label{createTaskRep-gas}
\end{figure*}

\item \textit{L2 vs L1 Performance:}

\color{black}

Now, let us discuss the results of the L2 versus L1 evaluation. Figure \ref{createTaskRep-gas} shows the gas consumption associated with the $createTask$ and $calculateNewRep$ functions in two different implementations. The transaction batching process in zkSync Rollup has three steps on L1: commit, prove, and execute, during which batches are committed, proven, and executed on L1. Each step incurs gas consumption, with the total gas being the sum of the gas expended in these three stages. The results clearly show that even when sending a single TX, the gas consumption is significantly reduced when passing through L2. Furthermore, these results also demonstrate that as the transaction complexity changes when calling the $calculateNewRep$ function, the gas cost does not change much and remains below that of the L1 execution. \\

\begin{table}[h]
\caption{Time overhead (s) for different functions}
\begin{center}
\begin{tabular}{|l|c|c|c|c|c|c|}
\hline
\bf Function calls (TX) & 1 &5&10&20&50& 100 \\ \hline  \hline

CreateTask & $0.13$ &$0.84$&$1.28$&$2.18$&$4.85$& $10.35$\\ 
SubmitSolution & $0.13$ &$0.81$&$1.28$&$2.26$&$4.61$& $9.9$\\ 
CalculateNewRep & $0.13$ &$0.83$&$1.29$&$2.19$&$5.09$& $9.94$\\ \hline

\end{tabular}
\label{tab:time-overhead}
\end{center}
\end{table}

\begin{table}[h]
\caption{Gas consumption of createTask: L1 vs L2}
\begin{center}
\begin{tabular}{|l|cccc|c|}
\hline
\bf Calls  &  \multicolumn{4}{c|}{\bf Dual layer (L2)} &  \multicolumn{1}{c|}{\bf Single Layer (L1)} \\ \hline  \hline
 --- &  \multicolumn{1}{c|}{commit.} &  \multicolumn{1}{c|}{verif.} &  \multicolumn{1}{c|}{exec.}  &  \multicolumn{1}{c|}{Total} &  \multicolumn{1}{c|}{Total}  \\ \hline 

1  &  \multicolumn{1}{c|}{38828} &  \multicolumn{1}{c|}{27260} &  \multicolumn{1}{c|}{23964}  &  \multicolumn{1}{c|}{\bf 90052}  &  \multicolumn{1}{c|}{\bf 212615}  \\ \hline

2  &  \multicolumn{1}{c|}{33964} &  \multicolumn{1}{c|}{27272} &  \multicolumn{1}{c|}{23964}  &  \multicolumn{1}{c|}{\bf 85200}  &  \multicolumn{1}{c|}{\bf 396636}  \\ \hline

5  &  \multicolumn{1}{c|}{33348} &  \multicolumn{1}{c|}{27284} &  \multicolumn{1}{c|}{23964}  &  \multicolumn{1}{c|}{\bf 84596}  &  \multicolumn{1}{c|}{\bf 896100}  \\ \hline

10  &  \multicolumn{1}{c|}{34348} &  \multicolumn{1}{c|}{27272} &  \multicolumn{1}{c|}{23952}  &  \multicolumn{1}{c|}{\bf 85572}  &  \multicolumn{1}{c|}{\bf 1792080}  \\ \hline

15  &  \multicolumn{1}{c|}{34324} &  \multicolumn{1}{c|}{27272} &  \multicolumn{1}{c|}{23964}  &  \multicolumn{1}{c|}{\bf 85560}  &  \multicolumn{1}{c|}{\bf 2646120}  \\ \hline

20  &  \multicolumn{1}{c|}{35396} &  \multicolumn{1}{c|}{27284} &  \multicolumn{1}{c|}{23964}  &  \multicolumn{1}{c|}{\bf 86644}  &  \multicolumn{1}{c|}{\bf 3966360}  \\ \hline

25  &  \multicolumn{1}{c|}{68744} &  \multicolumn{1}{c|}{29904} &  \multicolumn{1}{c|}{26584}  &  \multicolumn{1}{c|}{\bf 125232}  &  \multicolumn{1}{c|}{\bf 4412700}  \\ \hline

\end{tabular}
\label{tab:gas-layers}
\end{center}
\end{table}

\quad In Table \ref{tab:time-overhead}, we present the comprehensive end-to-end latency for the $createTask$, $submitSolution$, and $calculateNewRep$ functions. The results indicate that simultaneous computation for multiple transactions takes no more than a few seconds. We must mention here that within this duration, the transactions are included in L2 blocks.

\quad Table \ref{tab:gas-layers} illustrates the gas cost dynamics associated with multiple function calls simultaneously in both L1 and L2 (zkRollup) environments. In L1, the gas cost increases linearly with the number of calls. Since each call has a fixed gas cost, the resulting overall cost is equivalent to the cost of a single call multiplied by the number of calls. In L2 (zkRollup), on the other hand, the gas cost remains stable for up to 20 function calls, proving the effectiveness of the batching scheme within the zkRollup. The initial constant cost signifies the aggregation of up to 20 transactions into a single batch, significantly reducing gas expenses. Upon exceeding 20 function calls, a doubling of commit gas costs occurs, indicating the submission of a new batch to L1. Compared to costs obtained in the L1 scenario, there is a significant reduction of about 20X, demonstrating the consistent benefit of batching with zkRollups.

\end{enumerate}

\quad Overall, thanks to the integration of the zkRollup layer, the scalability of the entire system is improved and the gas cost of each transaction is considerably reduced. As a result, we are convinced that the framework we propose can effectively manage reputation and crowdsourcing tasks simultaneously.  

\section{Conclusion}
\label{sec:conclusion}
\quad In this paper, we presented RollupTheCrowd, an innovative blockchain-based crowdsourcing framework with a privacy-preserving reputation model and a L2 scaling solution. The use of zkRollups as a L2 solution enhances the scalability of the entire system and enables simultaneous management of reputation and crowdsourcing operations. The proposed framework incorporates an efficient, privacy-friendly reputation model. The designed model evaluates the trustworthiness of participants based on their crowdsourcing interactions. To reduce the load on our blockchain, we implement an off-chain storage scheme, improving the overall performance of RollupTheCrowd. The proof-of-concept we have provided supports the feasibility of our framework and the obtained results affirm its scalability and efficiency.

\quad In the future, our research will focus on anonymity aspects of blockchain-based reputation systems, to improve privacy protection while maintaining greater scalability.

\section*{acknowledgement}
This work was supported by the 5G-INSIGHT bilateral ANR-FNR project (ID: 14891397) / (ANR-20-CE25-0015-16), funded by the Luxembourg National Research Fund (FNR) and by the French National Research Agency (ANR), the Nouvelle-Aquitaine Region - B4IoT project, and The MIRES federation.

\end{document}